\documentclass[
 reprint,
 amsmath,amssymb,
 aps,
prb,
]{revtex4-2}

\usepackage{subfiles}
\usepackage{graphicx}
\usepackage{dcolumn}
\usepackage{bm}
\usepackage{hyperref}
\usepackage[mathlines]{lineno}
\usepackage{pifont}
\usepackage{subcaption}
\usepackage[dvipsnames]{xcolor}

\begin{document}

\preprint{APS/123-QED}

\title{Scaling behavior and giant field-enhancement of the thermal conductivity in the honeycomb antiferromagnet BaCo$_2$(AsO$_4$)$_2$} 

\author{Jiayi Hu\textsuperscript{1}}
\author{Ruidan Zhong\textsuperscript{2,*}}
\author{Peter Czajka\textsuperscript{1}}
\author{Tong Gao\textsuperscript{1,**}}
\author{R. J Cava\textsuperscript{2}}
\author{N. P. Ong\textsuperscript{1}}

\affiliation{{\normalfont\textsuperscript{1}}Department of Physics, Princeton University, Princeton, NJ, 08544, USA\\
    {\normalfont\textsuperscript{2}}Department of Chemistry, Princeton University, Princeton, NJ, 08544, USA\\
}

\date{\today}

\begin{abstract}
 The layered honeycomb material BaCo$_2$(AsO$_4$)$_2$ (BCAO) is of topical interest because its magnetic state is related to that of the Kitaev magnet $\alpha$-RuCl$_3$. Using thermal transport to probe how magnetic excitations interact with phonons in the magnetically disordered regime, we have uncovered an unusually large enhancement of the thermal conductivity $\kappa_{xx}$ in an in-plane magnetic field ${\bf H}$. Just above the N\'{e}el temperature $T_{\rm N}$, a field of 13 T increases $\kappa_{xx}$ by a factor $\sim 211$, much larger than reported previously in any magnetic insulator. Interestingly, $\kappa_{xx}(H,T)$ exhibits a scaling behavior in the entire magnetically disordered region that surrounds the ordered zigzag state. The ratio $\Delta \kappa_{xx}(H,T)/\kappa_{xx}(13,T)$, measured throughout the disordered region, collapses to a one-parameter scaling function ${\rm exp}(-1/gx)$ (where $x = \mu_{\rm B}B/k_{\rm B}T$ and $g$ is a constant). 
 
 
\end{abstract}
\maketitle

\section{Introduction}

The family of layered honeycomb spin-$\frac12$ magnetic materials has attracted considerable interest. In particular, the ruthenate $\alpha$-RuCl$_3$ with local moments derived from $4d$ orbitals in Ru has been identified~\cite{Banerjee,Jackelli} as the closest proximate to the Kitaev Hamiltonian ${\cal H}_K$~\cite{kitaev}. The low-temperature thermal transport properties have played a central role in illuminating the extent to which predictions based on ${\cal H}_K$ agree with experiment~\cite{Leahy,rucl3osc,matsuda,rucl3pthe,liuKhaliullin}. Interest has extended to related layered, honeycomb spin-$\frac12$ materials. In the cobaltate BaCo$_2$(AsO$_4$)$_2$ (BCAO), the local moments based on Co-$3d$ orbitals undergo a sharp transition to a zig-zag state at the N\'{e}el temperature $T_{\rm N}$ = 5.4 K~\cite{neutron1977,regnault1978,neutron2018}. In its magnetic, thermal and crystallographic properties, BCAO shares many features with $\alpha$-RuCl$_3$. A number of recent reports have compared their magnetic and transport properties~\cite{liuKhaliullin,tong,collin,shiyan,winter2022,abinitio,xymodel,pressure,intermediate2024,THz}. 

In the case of BCAO, the theoretical and experimental investigations appear to favor a $J_1$-$J_3$, $XXZ$ Hamiltonian instead of the Kitaev Hamiltonian~\cite{abinitio,collin,winter2022,xymodel,pressure,intermediate2024}. Below $T_{\rm N}$, the spins in BCAO exhibit a helical magnetic texture in the $ab$-plane. The spins, nominally aligned along the $b$-axis, exhibit a spiral structure with axis parallel to ${\bf a}^*$. In an in-plane magnetic field $\bf H$, the spins are polarized in a ferrimagnetic, colinear pattern above a field $H_{\rm c1}\sim 0.2$ T. Long-range magnetic order is suppressed at the critical field $H_{\rm c2}= 0.5$ T~\cite{neutron1977}\cite{neutron2018}\cite{collin}. The transition at the lower field $H_{\rm c1}$ has been investigated by magnetization and terahertz spectroscopy experiments~\cite{tong,collin,shiyan}. A recent study combining angular dependence of magnetization and dilatometry suggests an intermediate phase appears before the spins are fully polarized~\cite{intermediate2024}. Another thermal transport experiment measured down to subKelvin temperatures supports the KQSL picture, yet other theoretical proposal attribute it to a Dirac spin liquid phase~\cite{shiyan,bose2023proximate}.

Our work here focusses on the magnetically disordered regime which covers a large fraction of the $H$-$T$ plane. We report an unusual scaling behavior of the thermal conductivity $\kappa_{xx}(H,T)$ in BCAO observed throughout the magnetically disordered region below 10 K. In zero $H$, strong inelastic scattering of phonons from spin excitations severely suppresses phonon conduction. In an in-plane magnetic field $\bf H$, $\kappa_{xx}(H,T)$ increases dramatically following a one-parameter scaling function. Just above $T_{\rm N}$, the field-induced enhancement factor ($>200$) exceeds that previously reported in all magnetic insulators to our knowledge. 

Underlying the giant field-enhancement of $\kappa_{xx}$ is the intense scattering of phonons by the spins which leads to  ``localization'' of the phonons in zero $H$. Our results suggest that the non-analytic behavior of $\kappa_{xx}(H,T)$ in the limit $H\to 0$, as well as the scaling behavior over a large fraction of the magnetically disordered region, may well be signature hallmarks of strong spin-phonon coupling.

In several topical areas, e.g. spin qubits based on NV centers in diamond~\cite{Wolfowicz}, spin-phonon coupling is a key topic of interest. The scaling behavior uncovered here suggests an incisive way to probe these couplings in magnetic materials. 

\begin{figure*}
    \centering
    \includegraphics[width=\textwidth]{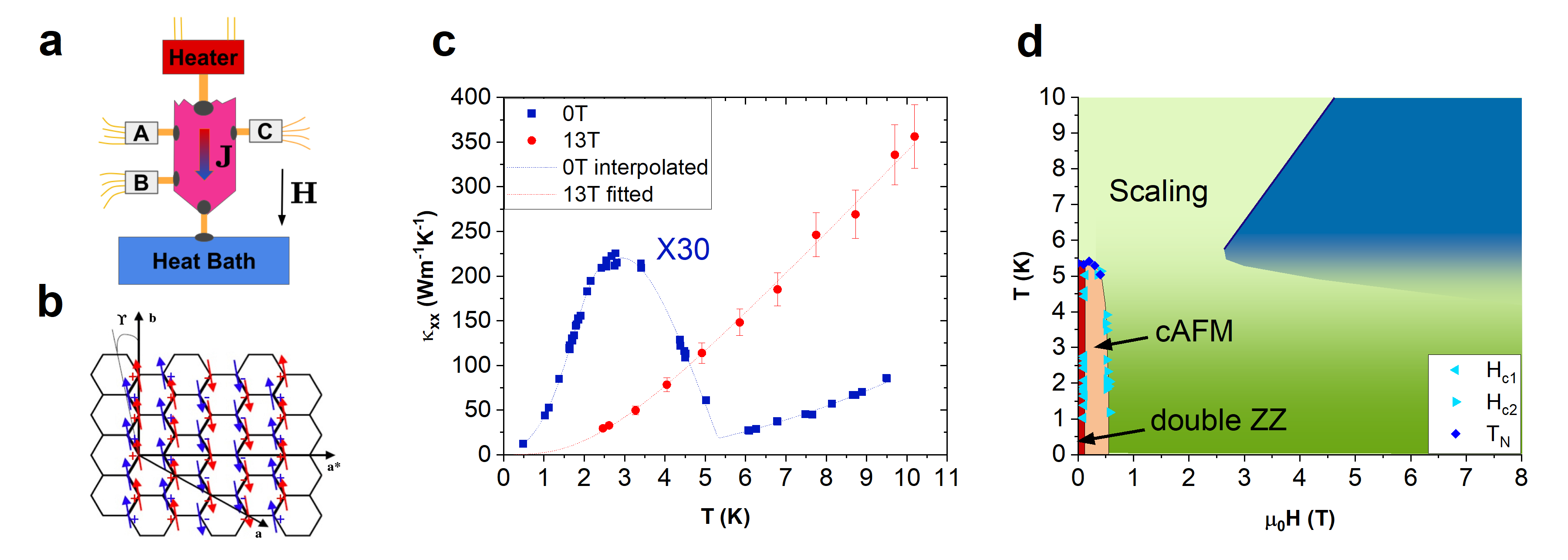}
    \caption{
    Panel (a): Schematic of the crystal attached to the heater, 3 thermometers and the bath by gold wires. The applied $\bf H$ direction is indicated. 
    Panel (b): The two-up, two-down zigzag phase inferred from neutron scattering~\cite{neutron2018}.
    Panel (c): Comparison of the zero-$H$ $\kappa(T)$ (square blue symbols, multiplied by a factor of 30) with $\kappa_{xx}$ vs. $T$ measured in an in-plane 13-T magnetic field (red symbols). The faint blue curve passing through the blue symbols is an interpolation supplemented by measurements in zero $H$ in which $T$ is continuously varied. The red curve is a fit to the 13-T data using the standard Callaway expression. More details can be found in the Supplement~\cite{Supplement}.
    Panel (d): The inferred phase diagram of BCAO in the $H$-$T$ plane. The magnetically ordered zigzag state is shaded red and coral. Throughout the disordered region (shaded green), $\kappa_{xx}(H,T)$ obeys a one-parameter scaling behavior. Deviation from this scaling occurs in the blue region. The fields $H_{\rm c1}$ and $H_{\rm c2}$ are obtained from data presented in Figure \ref{Profiles}a.
    }
    \label{Setup}
\end{figure*}

\section{Experiment}

To measure the thermal conductivity tensor $\kappa_{ij}$ of the BCAO crystal, we heat sunk 3 RuO\textsubscript{2} thermometers (matched and calibrated) to its edges using stycast and gold wires (Fig. \ref{Setup}a). One short edge of the crystal is heat sunk to the bath (brass substrate) while the opposite edge is attached to a heater via gold wires. The sample space is kept at a pressure less than $5\times 10^{-5}$ mbar. We label as $\bf \hat{x}$ the common axis of the heat current density $\bf{J}$ and field $\bf H$. 

Typically, we measure the field profile of $\kappa_{xx}(H,T)$ keeping the bath temperature $T_{\rm bath}$ regulated at a set point as $H$ is swept from 0 to 13 T and back. At low $T$, the applied heat current invariably leads to a finite shift $\delta T = T-T_{\rm bath}$ between the sample's temperature $T$ and $T_{\rm bath}$ (the shift is usually nearly $H$ independent). Here, however, the anomalously large (211-fold) increase in $\kappa_{xx}(H,T)$ causes $\delta T(H,T)$ to be strongly field dependent. In all field profiles reported here, we have recorded $\delta T(H,T)$ to correct for these shifts. We also adopted steps to avoid errors arising from large magnetocaloric effects which lead to long relaxation times for both the sample and bath to reach equilibrium whenever $H$ undergoes a step-change (Secs. 1, 3 and 4 of Supplement~\cite{Supplement}). Finally, the tendency of the crystal to fracture under strain exerted by magnetic torque restricts measurements so far to in-plane $\bf H$.

Motivated by the observation of an unusual planar thermal Hall conductivity $\kappa_{xy}$ in the Kitaev magnet $\alpha$-RuCl$_3$~\cite{matsuda,rucl3pthe}, we have searched for a planar thermal Hall signal in BCAO focusing on the disordered region close to the ordered phase. Despite a careful search we did not detect a finite $\kappa_{xy}$ signal in BCAO within our resolution (See Sec. 7 of Supplement~\cite{Supplement}).

\section{Field and temperature profiles of thermal conductivity}

The thermal conductivity $\kappa_{xx}(H,T)$ of BCAO displays a rich set of features as a function of temperature $T$ and an in-plane magnetic field ${\bf H}$. In the absence of $\bf H$, $\kappa(T)$ ($\equiv \kappa_{xx}(0,T)$, plotted as blue symbols in Fig \ref{Setup}c) slowly decreases to an anomalously low value ($\sim 0.6$ W/mK) as $T$ is lowered to 5.4 K. At the N\'{e}el temperature $T_{\rm N}$ (5.4 K), the spins undergo a phase transition to an antiferromagnetic state with zig-zag order. In the ordered state, $\kappa$ rises to a broad peak that reflects contributions from phonons and magnons in the ordered state.  

In our experiment, we observe that long-range order is abruptly suppressed by an in-plane field $H\sim$ 0.5 T, consistent with previous studies~\cite{tong,THz,pressure,collin}. As described below, phonons remain strongly coupled to spin excitations throughout the disordered region surrounding the ordered state (green region in Fig. \ref{Setup}d). The strong coupling is steadily weakened as $H$ increases. Eventually, at large $H$ (13 T), scattering of phonons by the spins becomes negligible, and $\kappa_{xx}$ attains large values (red symbols in Panel c). The field enhancement, expressed by the ratio $\kappa_{xx}(H,T)/\kappa(T)$ ($\sim 211$ just above $T_{\rm N}$) exceeds that in all other magnetic insulator reported to date. Our key result is that, over the entire magnetically disordered region (green region in Panel d), $\kappa_{xx}(H,T)$ obeys a one-parameter scaling function.

\begin{figure*}
    \centering
    \includegraphics[width=1\textwidth]{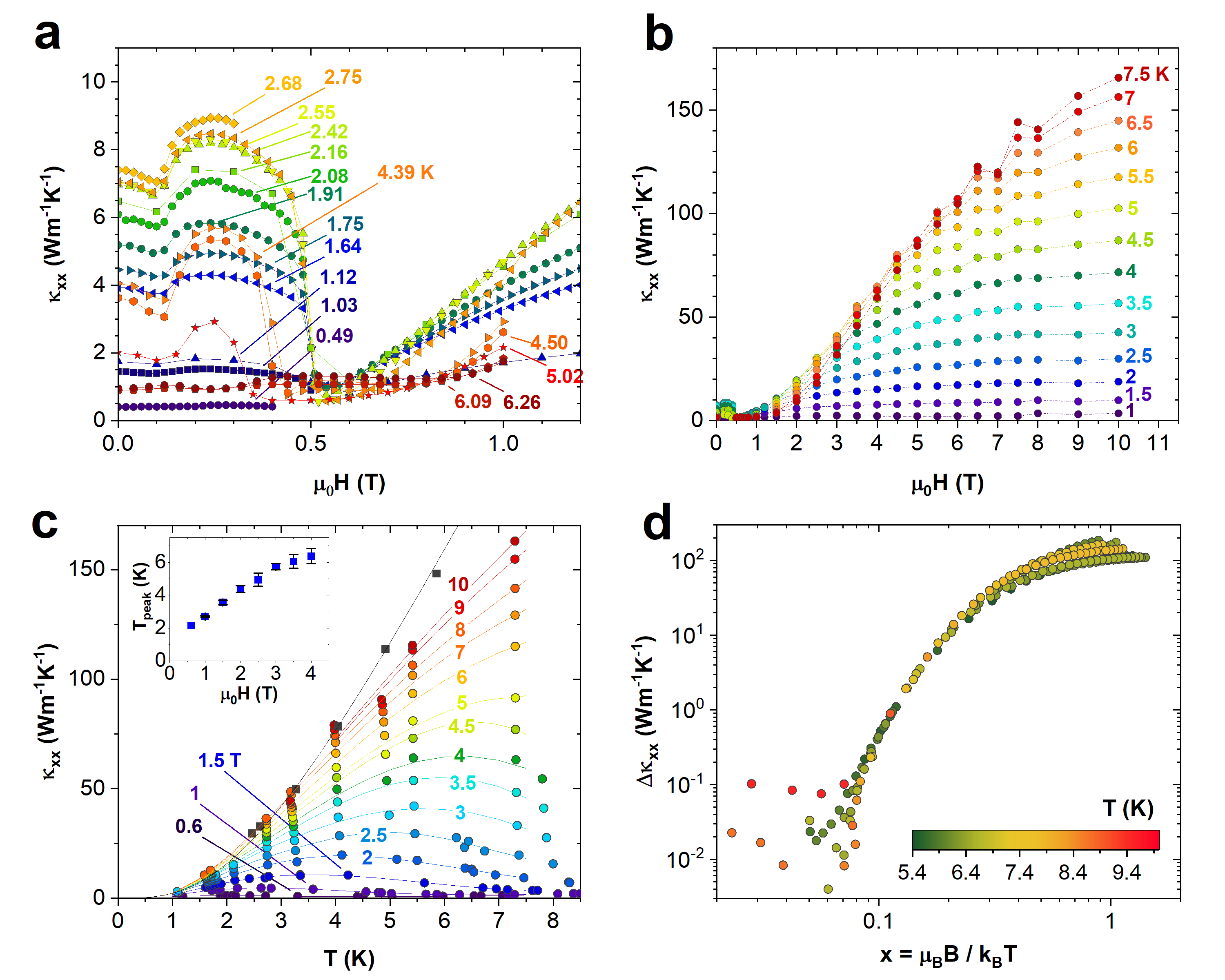}
    \caption{Profiles of $\kappa_{xx}$ vs. field and temperature. Panel (a): Field profiles of $\kappa_{xx}(H,T)$ at low $H$ ($\mu_0H< 1.2$ T). In each curve, $T$ is fixed at the values indicated. Curves in the temperature interval $1.6 < T< 2.8$ K display a vertical cliff that signals the abrupt collapse of long-range magnetic order as $H$ approaches $H_{\rm c2}$ = 0.5 T. At low $H$, a second sharp feature $H_{\rm c1}$ ($\sim$ 0.1 T) is observed. 
    Panel (b): Field profiles of $\kappa_{xx}$ extended to 11 T, with $T$ fixed at the values indicated. At each $T$, $\kappa_{xx}$ initially increases exponentially with $H$. At large $H$, it gradually saturates to its high-field limit. Profiles shown in Panel (a) are visible as the small mesa below 0.5 T. 
     Panel (c): Curves of $\kappa_{xx}(H,T)$ plotted versus $T$ with $H$ fixed at the values indicated. The curve $\kappa_{xx}(13,T)$ at the maximum applied $H$ (13 T) serves as an envelope that caps the maximum value attained at each temperature $T$. Each curve (e.g. the one at $H$ = 4 T) initially tracks closely $\kappa_{xx}(13,T)$ as $T$ is raised above 1 K, but deviates away from $\kappa_{xx}(13,T)$ as $T$ is raised (at $\sim 2.5$ K). Curves taken at larger $H$ (e.g. 10 T) deviate from $\kappa_{xx}(13,T)$ at a higher temperature (4 K). This pattern reflects the one-parameter scaling behavior discussed below. Note the anomalously low values of $\kappa_{xx}$ in the curve at 0.6 T. 
 Panel (d): First hints of scaling behavior above $T_N$. When we plot the field-induced change $\Delta\kappa_{xx}(H,T)$ vs. the scaling variable $\mu_0H/T$, we find that all data points above $T_N$ converge to the same curve when the scaling variable $\mu_0H/T$ falls below $\sim 0.4$ T/K. The temperature of each data point is given by the color scale.
    }
    \label{Profiles}
\end{figure*}

Figure \ref{Profiles}a displays field profiles of $\kappa_{xx}$ within the ordered state with $T$ fixed at the values indicated. At the lowest $T$ (0.49 K), the profile of $\kappa_{xx}$ is virtually flat. When $T$ is raised (1.64 to 2.75 K) the field profiles feature a dome that is abruptly truncated on the large-$H$ side by a vertical cliff as $H$ approaches $H_{\rm c2}$ (0.5 T), and long-range order is suppressed. On the low $H$ side, the dome displays a notch that corresponds to the lower critical field $H_{\rm c1}$ ($\sim 0.12$ T). In previous reports, $H_{\rm c1}$ was interpreted as a transition between either an incommensurate spiral state or a double zigzag pattern to the cAFM state~\cite{neutron2018,THz}. Unlike in Refs. ~\cite{tong,shiyan}, we observe only a single sharp transition within the field interval 0.1--0.2 T. In the disordered phase above $H_{\rm c2}$, $\kappa_{xx}$ increases steeply with $H$. This is a small fraction of the scaling pattern uncovered below.

Figure \ref{Profiles}b shows the field profiles of $\kappa_{xx}$ with the $H$ axis extended to 11 T (the profiles in the ordered phase are now reduced to small mesas visible below 0.5 T). At each $T$ (e.g. 3 K), $\kappa_{xx}$ increases from a small value ($\sim$ 1 W/mK) before slowly saturating to a maximum value ($\sim 42$ W/mK) as $H$ increases to 13 T. All the profiles appear to follow the same functional form, with a pattern amenable to simple analysis.

To this end, we turn to the orthogonal cuts in which $\kappa_{xx}(H,T)$ is plotted vs. $T$ with $H$ fixed (Fig. \ref{Profiles}c). At each fixed value of $H$, the curve of $\kappa_{xx}$ vs. $T$ is now non-monotonic, exhibiting a broad maximum at the temperature $T_{\rm p}(H)$. As shown in the inset, $T_{\rm p}(H)$ increases roughly linearly with $H$. This suggests that, in the disordered region, $\kappa_{xx}$ depends on $H$ solely through the scaling variable 
\begin{equation}
    x = \mu_{\rm B}B/k_{\rm B}T,
\label{x}
\end{equation}
where $\mu_{\rm B}$ and $k_{\rm B}$ are, respectively, the Bohr magneton and the Boltzmann constant, and $B = \mu_0H$ is the magentic flux density with $\mu_0$ the vacuum permeability.

To test this preliminary hint of scaling behavior, we plot in Fig. \ref{Profiles}d the field-induced change $\Delta\kappa_{xx}(H,T) = \kappa_{xx}(H,T) - \kappa_{xx}(0,T)$ versus $x$ in $\log$-$\log$ scale for $T$ between 5.4 and 9.5 K. We find that the data merge to define a scaling curve that extends over nearly 3 decades in $\Delta\kappa_{xx}(H,T)$. 

Another important feature in Fig. \ref{Profiles}c is the role of $\kappa_{xx}(13,T)$ (black squares) relative to $\kappa_{xx}$ measured at lower fields throughout the disordered region. Combining the perspectives afforded by the fixed-$T$ and fixed-$H$ cuts, we infer that when $T$ is fixed (e.g. at 4 K), $\kappa_{xx}$ increases towards $\kappa_{xx}(13,T)$ but does not exceed it. 

Physically, this implies that, in the high-field limit, phonon conduction asymptotes to the regime in which $\ell$, restricted by scattering from crystalline disorder only, becomes independent of $T$. In this limit, the $T$ dependence observed in $\kappa_{xx}(H,T)= \frac13 c_{gs}(T)v_s\ell$ arises from the heat capacity $c_{gs}(T)$ of the spin-phonon system ($v_s$ is the group velocity of the acoustic branch). As inferred from the field profiles in Fig. \ref{Profiles}b, $\kappa_{xx}(13,T)$ is a good approximation to this high-field limit below $T$ = 6 K. The approximation is less satisfactory above 6 K as saturation is approached at increasingly large $H$. The dashed curve in Fig. \ref{Profiles}c is a fit to $\kappa_{xx}(13,T)$ using the Callaway model in this limit (see Sec. 2 of Supplement~\cite{Supplement}).

\section{Scaling behavior}
\begin{figure*}
    \centering
    \includegraphics[width=1\textwidth]{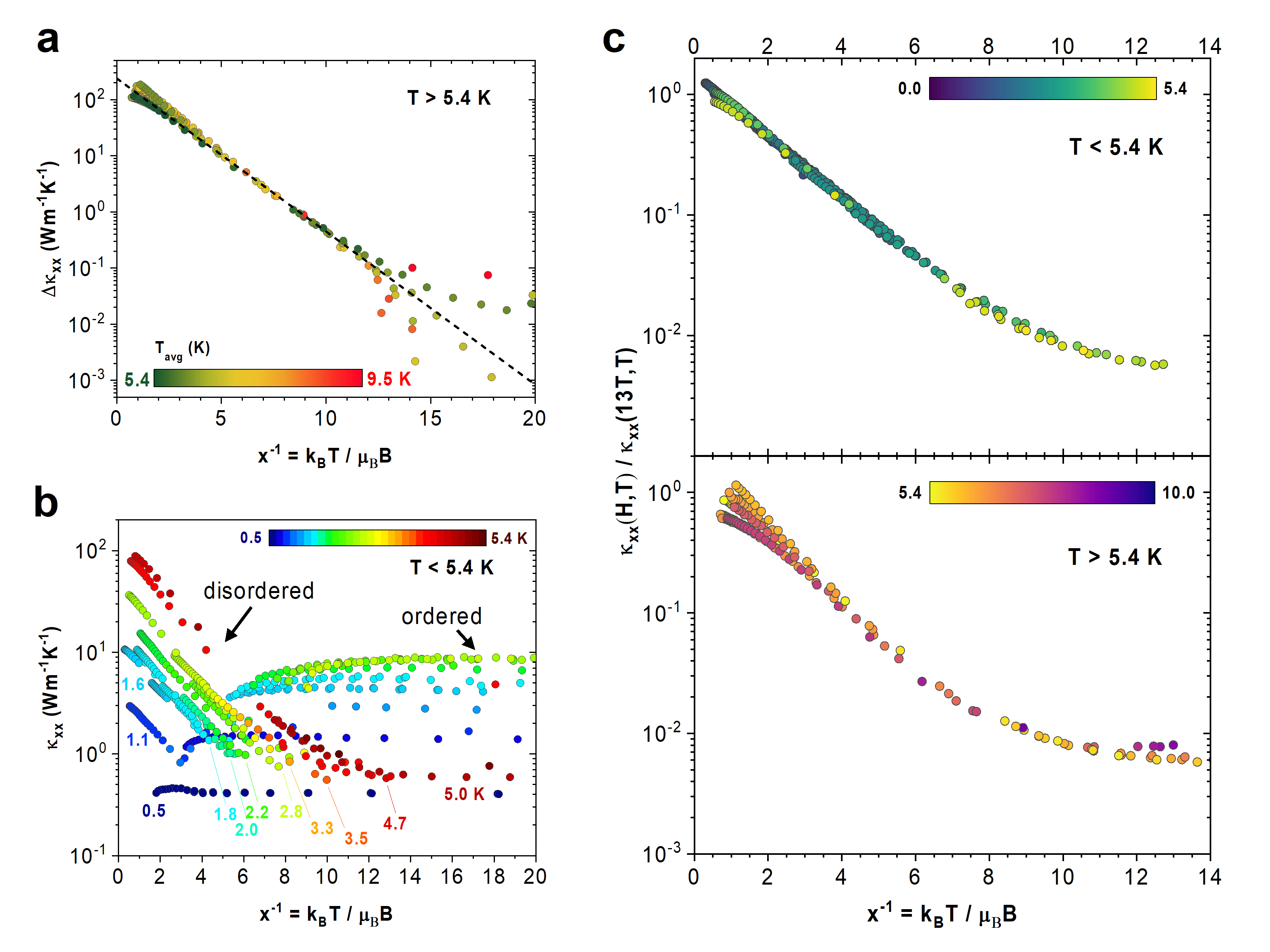}
    \caption{Uncovering the scaling behavior.
    (Panel a) The analytical form of the scaling behavior implied in Fig. \ref{Profiles}d becomes apparent in a semilog plot of $\Delta\kappa_{xx}$ vs. $1/x$ with $x$ defined in Eq. \ref{x}. All data points taken above $T_N$ fall on the dashed line given by Eqs. \ref{DKeqn} and \ref{x}. The equation of the fit (dashed line) is ln\( \,\Delta\kappa_{xx}=-0.628\,x^{-1}+5.476\), which gives $g=1.6$. The scatter as $x\to 0$ reflects uncertainties in $\Delta\kappa_{xx}$. 
    (Panel b) Finding the missing pre-factor in the scaling form. At temperatures below 5.4 K, the data points $\kappa_{xx}(H,T)$ segregate into two distinct groups when plotted vs. $1/x$. In the ordered phase ($H<H_{\rm c2}$), the data follow nearly flat trajectories. By contrast, data in the disordered region ($H>H_{\rm c2}$) fall on diagonal curves that match the dashed line in Panel (a), except that they are systematically displaced downwards as $T\to 0$. This suggests that $\kappa_{xx}$ should be divided by a $T$-dependent factor $T^\beta$, which we approximate by the measured curve $\kappa_{xx}(13,T)$.  
    (Panel c): Semilog plots of $\kappa_{xx}/\kappa_{xx}(13,T)$ vs. $1/x$ for $T<$ 5.4 K (upper panel) and $T>$ 5.4 K (lower). Once $\kappa_{xx}$ is divided by $\kappa_{xx}(13,T)$, the diagonal curves in Panel (b) collapse to a single curve, which represents the universal curve given in Eq. \ref{DKeqn}. The data above 5.4 K also collapse onto the same curve (lower panel). Hence, over the disordered region (green region in Fig. \ref{Setup}d), scaling behavior prevails. Departure from scaling becomes observable when $1/x$ decreases below $\sim 3$ (dark blue region in Fig. \ref{Setup}d).
    }
    \label{Scaling}
\end{figure*}

The functional form of the scaling curve in Fig. \ref{Profiles}d becomes evident when we plot $\Delta\kappa_{xx}$ vs. $1/x$ in semilog plot. As shown in Fig. \ref{Scaling}a, all the data above 5.4 K collapse onto the straight line corresponding to the function $\Delta\kappa_{xx}\sim {\rm e}^{-1/gx}$ with $g$ a constant.

The simple scaling based on $x$ accounts only for the $H$ dependence of $\Delta\kappa_{xx}(T,H)$. The further dependence on $T$ requires more care. This becomes apparent if we attempt to apply the same analysis to the data below 5.4 K in the disordered region ($H>H_{\rm c2}$).

The semilog plot of $\kappa_{xx}$ vs. $1/x$ shown in Fig. \ref{Scaling}b reveals that the data segregate into two distinct groups. Data in the ordered phase (not relevant to scaling) follow flat trajectories that are nearly independent of $x$. By contrast, the data in the disordered region fall along diagonal curves that mimic the scaling function except that they are now displaced vertically by a factor of the form $T^\beta$ with exponent $2<\beta<3$, which we infer arises from the heat capacity $c_{gs}(T)$. From the discussion on Fig. \ref{Profiles}c, we anticipate that the offsets can be cancelled if we divide $\kappa_{xx}(H,T)$ by $\kappa_{xx}(13,T)$.

This inference is confirmed when we replot $\kappa_{xx}(H,T)/\kappa_{xx}(13,T)$ vs. $1/x$ in Fig. \ref{Scaling}c (upper panel). Now all data throughout the disordered region below 5.4 K collapse onto one scaling curve. A similar plot shows that data above 5.4 K also collapse onto the same curve (lower panel). As noted above, deviation from scaling appears when $x$ exceeds $\sim 0.3$.

Hence, throughout the disordered region below and above 5.4 K, the ratio $\Delta\kappa_{xx}(H,T)/\kappa_{xx}(13,T)$ is described by the one-parameter scaling form 
\begin{equation}
\frac{\Delta\kappa_{xx}(H,T)}{\kappa_{xx}(13,T)} = A\exp[-\frac{1}{gx}]
\label{DKeqn}
\end{equation}
where $x$ is given by Eq. \ref{x} and $A$ and $g$ are constants.

The function ${\rm e}^{-1/gx}$ implies that $\Delta\kappa_{xx}$ initially increases from zero at a very slow rate (the non-analytic limit $H\to 0$ requires all derivatives to vanish). As discussed, the ultra-slow initial growth may be seen in Fig. 2a. However, once $B$ rises above 1 Tesla, $\Delta\kappa_{xx}$ undergoes an exponential increase before ultimately saturating to a very large value consistent with phonons in a crystal with a low density of lattice and spin disorder.

We sketch the physical picture suggested by our results. The single-parameter scaling form constrains the candidate scenarios for describing the field-enhanced $\kappa_{xx}$. 

First, the anomalously low value of $\kappa(T)$ in the fluctuating interval $5.4 < T< 10$ K at $H = 0$ implies strong spin-phonon coupling which leads to repeated absorption or emission of phonons accompanied by spin flips. 
Secondly, the non-analytic nature in Eq. \ref{DKeqn} in the limit $x\to 0$ implies that, in weak $H$, deviation of $\Delta\kappa_{xx}(H,T)$ from zero is initially nearly unresolvable until $H$ exceeds a field scale of $\sim 1$ T. Then $\kappa_{xx}$ increases exponentially. This strongly non-linear increase is not amenable to a perturbative treatment. 
(We also point out that Eq. \ref{DKeqn} is very different from the familiar expression ${\rm e}^{-\Delta_{\rm b}/k_{\rm B}T}$ describing thermal activation of charge carriers across a band gap $\Delta_{\rm b}$.)  

Throughout the disordered region, we infer that the phonons and spin excitations constitute a tightly coupled system in which $\ell_0$ (the phonon mean-free path at $H=0$) is limited to less than 30 nm by intense scattering. Although the scattering may be elastic or inelastic, the scaling form implies that inelastic processes are dominant in the disordered region. By analogy with experiments on the honeycomb system $\alpha$-RuCl$_3$~\cite{Ponomaryov2017,Ponomaryov2020,Wellm,Balz} we may expect that spin excitations in the disordered state exhibit a broad, featureless spectrum that is gapless in zero $H$. The low-lying spin excitations interact strongly with phonons at a select population of sites, which absorb and emit phonons. At temperature $T$, the phonon energy distribution peaks sharply at $\hbar\omega \simeq k_{\rm B}T$.

An applied Zeeman field opens a gap $\Delta = g\mu_{\rm B}H$ in the spin spectrum where $g$ is the effective gyromagnetic parameter for the low-lying spin excitations. Phonon annihilation is blocked when the gap exceeds $k_{\rm B}T$, i.e.
\begin{equation}
    g\mu_{\rm B} B > k_{\rm B}T.
    \label{escape}
\end{equation}
In effect, the identified absorption sites act as escape hatches that are propped open by $H$. Rendering these sites transparent to phonons allows them to escape the trapping well. Despite their small population, these sites play an outsize role in determining the initial growth of $\kappa_{xx}(H,T)$.

When $H$ is weak, the average spacing between escape hatches far exceeds $\ell_0$. Each liberated phonon does not travel far before becoming retrapped. 
Hence the initial increase in $\kappa_{xx}$ is nearly unresolvable, as shown in Figures 2a and 2b, and implied by the non-analyticity of ${\rm e}^{-T/B}$ in the limit $B\to 0$. 
Crucially, by Eq. \ref{escape}, the field at which phonons escape at the identified sites is proportional to $T$. Hence, the field-induced increase in $\ell$ automatically satisfies the scaling in Eq. \ref{DKeqn}.

Increasing $H$ weakens the amplitude of the fluctuating magnetic order by increasing the polarization of spins, which causes the population of escape hatches to proliferate. In analogy with percolation conductance, the largest enhancement of $\ell$ occurs when connected paths form skeletal spines that span the crystal.
When $H$ exceeds $\sim 1$ T, rapid proliferation of the hatches enables the connectivity of phonon propagation paths to exceed the percolation limit, and $\kappa_{xx}$ undergoes the exponential growth observed. 

\section{Comparison with other magnetic insulators}
Experiments showing large field-induced enhancement of $\kappa_{xx}$ have been reported in several layered magnetic insulators. We briefly survey the results beginning with the largest previously reported (Fig. \ref{Compare} and Table S1 in Supplement~\cite{Supplement}). Zhao \emph{et al.} found that, in Co$_3$V$_2$O$_8$, $\kappa_{xx}$ increases 100 fold in a field of 14 T~\cite{co3v2o8} at $7.5K$. They interpret the increase as caused by a series of magnetic phase transitions occuring in a narrow window (5 transitions between 6 and 12 K). While there are fewer known phase transitions in the case of BCAO, recent dilatometry experiment indeed suggests more intermediate phases~\cite{intermediate2024}. One difference we want to point out though is that BCAO has only one critical temperature, and it has been relatively robust even when applied pressure~\cite{pressure}. More importantly, the field-induced enhancement persists up to as high as 13T well beyond the critical field $H_{c2}=0.5T$, and the scaling behavior also covers a wide regime in the phase diagram surrounding the ordering states. As a result, a succession of phase transitions doesn't seem to be the whole story, though intense spin fluctuations do seem to be one of the key components.

A recent experiment on another honeycomb Kitaev magnet candidate Na$_2$Co$_2$TeO$_6$ found that $\kappa_{xx}$ increases by a factor of 70 in a field of 16 T~\cite{nctoHess}. The authors attribute this intense scattering to the abundance of magnetic excitations above a triple-Q ground state. The authors do not report any scaling behavior in the observed $\kappa_{xx}$ vs. $H$. 

In the Kitaev magnet $\alpha$-RuCl$_3$, $\kappa_{xx}$ has been reported to increase 13-fold compared to the value at $7T$, the critical field of its zig-zag state~\cite{rucl3resonance}. The ratio is smaller if compared to its zero field value (See Table 1 in the Supplement~\cite{Supplement}). The authors are able to fit $\kappa_{ab}$ with a modified Callaway model, adding a relaxation term that takes into account the resonance between phonon and magnetic gap, whose size is linear to the applied field. Motivated by this study, We also tried fitting our data at finite field with the modified Callaway model, with spin-phonon resonance terms. However, we have not succeeded in fitting our data with physically meaningful parameters (the exception is the curve at 13 T where $\ell$ is limited by disorder scattering). In fact, the non-analyticity inherent in the $H\to 0$ limit in Eq.\ref{escape} rules out a Callaway-type description in the scaling regime in BCAO. Because it is not possible to reproduce the non-analyticity in a Calloway-type formulation, our conclusion is that the empirical fits used in $\alpha$-RuCl$_3$ cannot be applied to BCAO.

\begin{figure}
    \centering
    \includegraphics[width=0.5\textwidth]{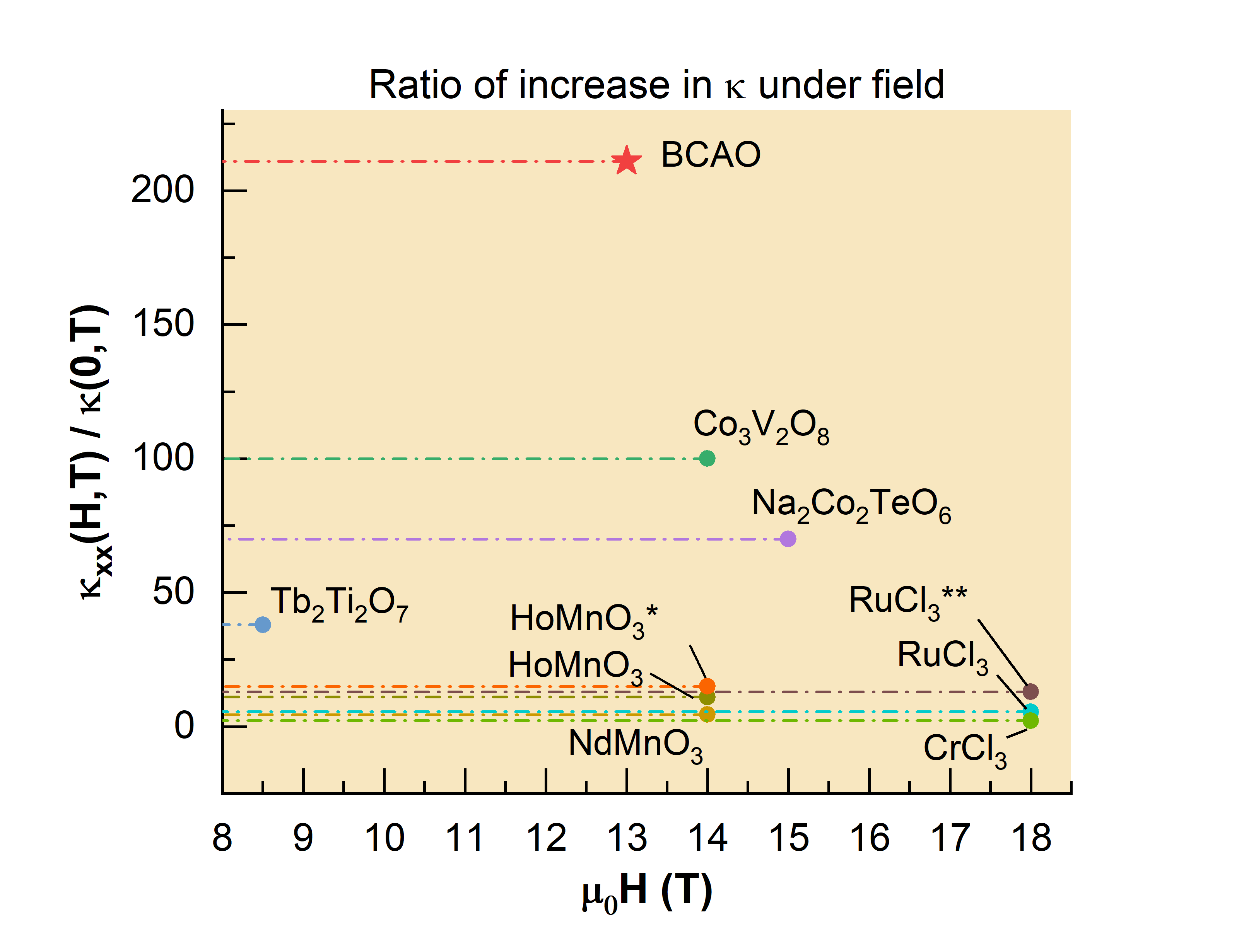}
    \caption{Comparison of the field enhancement of $\kappa_{xx}$ (expressed as a ratio) with the ratio reported for other magnetic insulators\cite{co3v2o8,nctoHess,tb2ti2o7,homno3,ndmno3,crcl3,rucl3resonance}. BCAO is marked with a red star. In the ratio for HoMnO$_3^*$ the denominator is $\kappa_{xx}$(2.2T). For RuCl$_3^{**}$, the denominator is $\kappa_{xx}$(7.5T). See Sec. 6 in Supplement~\cite{Supplement} for detailed information.
    }
    \label{Compare}
\end{figure}
\vspace{1cm}

\centerline{***************}

\newpage
\vspace{1cm}\noindent
*\emph{Present address of R.Z.}: Tsung-Dao Lee Institute and School of Physics and Astronomy, Shanghai Jiao Tong University, Shanghai 201210, China\\
**\emph{Present address of T.G.}: Building 3, West Area, Xibeiwang East Street, Haidian District, 
Beijing 100193, China\\

\vspace{3mm}\noindent
{\bf Acknowledgement}
The thermal transport experiments were supported by the US Department of Energy (Basic Energy Sciences) under contract DE-SC0017863 to N.P.O. An award from the U.S. National Science Foundation (DMR 2011750) supported the materials growth (R.J.C.) and acquisition of cryogenic equipment (N.P.O.). N.P.O. acknowledges support from the Gordon and Betty Moore Foundation EPiQS initiative (grant GBMF9466).

\vspace{3mm}
\noindent
{\bf Author contributions}\\
J.H., P.C. and N.P.O. conceptualized the experiment. J.H. performed all measurements with initial guidance from P.C. Data in an earlier experiment by T.G. provided substantial insights. The present data were analyzed by J.H. and N.P.O. The crystals were grown and characterized by R.Z. and R.J.C. The manuscript was written by J.H. and N.P.O. with input from all authors.

\vspace{3mm}
\noindent
{\bf Competing financial interests}\\
The authors declare no competing financial interests.

\vspace{3mm}
\noindent
{\bf Additional Information}\\
Supplementary Materials is available in the online version of the paper.

\vspace{3mm}
\noindent
{\bf Correspondence and requests for materials}
should be addressed to J.H. (jiayih@princeton.edu) or N.P.O. (npo@princeton.edu).


\vspace{3cm}



\begin{thebibliography}{99}

\bibitem{Banerjee}Banerjee, A., Bridges, C.A., Yan, J.Q., Aczel, A.A., Li, L., Stone, M.B., Granroth, G.E., Lumsden, M.D., Yiu, Y., Knolle, J. and Bhattacharjee, S., Proximate Kitaev quantum spin liquid behaviour in a honeycomb magnet. {\em Nature Materials}. \textbf{15}, 733-740 (2016).
\bibitem{Jackelli}Jackeli, G. and Khaliullin, G. Mott Insulators in the Strong Spin-Orbit Coupling Limit: From Heisenberg to a Quantum Compass and Kitaev Models. {\em Phys. Rev. Lett.}. \textbf{102}, 017205 (2009,1).
\bibitem{kitaev}Kitaev, A. Anyons in an exactly solved model and beyond. {\em Annals Of Physics}. \textbf{321}, 2-111 (2006), January Special Issue.
\bibitem{Leahy}Leahy, I., Pocs, C., Siegfried, P., Graf, D., Do, S., Choi, K., Normand, B. and Lee, M. Anomalous thermal conductivity and magnetic torque response in the honeycomb magnet $\alpha$-RuCl$_3$. {\em Physical Review Letters}. \textbf{118}, 187203 (2017).
\bibitem{rucl3osc}Czajka, P., Gao, T., Hirschberger, M., Lampen-Kelley, P., Banerjee, A., Yan, J., Mandrus, D., Nagler, S. and Ong, N. Oscillations of the thermal conductivity in the spin-liquid state of $\alpha$-RuCl$_3$. {\em Nature Physics}. \textbf{17}, 915-919 (2021).
\bibitem{matsuda}Kasahara, Y., Ohnishi, T., Mizukami, Y., Tanaka, O., Ma, S., Sugii, K., Kurita, N., Tanaka, H., Nasu, J., Motome, Y. and Others, Majorana quantization and half-integer thermal quantum Hall effect in a Kitaev spin liquid. {\em Nature}. \textbf{559}, 227-231 (2018).
\bibitem{rucl3pthe}Czajka, P., Gao, T., Hirschberger, M., Lampen-Kelley, P., Banerjee, A., Quirk, N., Mandrus, D., Nagler, S. and Ong, N. Planar thermal Hall effect of topological bosons in the Kitaev magnet $\alpha$-RuCl$_3$. {\em Nature Materials}. \textbf{22}, 36-41 (2023).
\bibitem{liuKhaliullin}Liu, H. and Khaliullin, G. Pseudospin exchange interactions in d$^7$ cobalt compounds: Possible realization of the Kitaev model. {\em Phys. Rev. B}. \textbf{97}, 014407 (2018,1).
\bibitem{neutron1977}Regnault, L., Burlet, P. and Rossat-Mignod, J. Magnetic ordering in a planar X - Y model: BaCo$_2$(AsO$_4$)$_2$. {\em Physica B+C}. \textbf{86-88} pp. 660-662 (1977).
\bibitem{regnault1978}Regnault, L., Rossat-Mignod, J., Villain, J. and Combarieu, A. Specific heat of the quasi 2D-XY helimagnets BaCo$_2$(AsO$_4$)$_2$ and BaCo$_2$(PO$_4$)$_2$. {\em Journal De Physique Colloques}. \textbf{39}, C6-759-C6-761 (1978)
\bibitem{neutron2018}Regnault, L., Boullier, C. and Lorenzo, J. Polarized-neutron investigation of magnetic ordering and spin dynamics in BaCo$_2$(AsO$_4$)$_2$ frustrated honeycomb-lattice magnet. {\em Heliyon}. \textbf{4}, e00507 (2018).
\bibitem{tong}Zhong, R., Gao, T., Ong, N. and Robert J. Cava Weak-field induced nonmagnetic state in a Co-based honeycomb. {\em Science Advances}. \textbf{6}, eaay6953 (2020).
\bibitem{collin}Halloran, T., Desrochers, F., Zhang, E., Chen, T., Chern, L., Xu, Z., Winn, B., Graves-Brook, M., Stone, M., Kolesnikov, A. and Others, Geometrical frustration versus Kitaev interactions in BaCo$_2$(AsO$_4$)$_2$. {\em Proceedings Of The National Academy Of Sciences}. \textbf{120}, e2215509119 (2023).

\bibitem{abinitio}Maksimov, P., Ushakov, A., Pchelkina, Z., Li, Y., Winter, S. and Streltsov, S. Ab initio guided minimal model for the “Kitaev” material BaCo$_2$(AsO$_4$)$_2$: Importance of direct hopping, third-neighbor exchange, and quantum fluctuations. {\em Physical Review B}. \textbf{106}, 165131 (2022)
\bibitem{pressure}Huyan, S., Schmidt, J., Gati, E., Zhong, R., Cava, R., Canfield, P. and Bud'Ko, S. Hydrostatic pressure effect on the Co-based honeycomb magnet BaCo$_2$(AsO$_4$)$_2$. {\em Physical Review B}. \textbf{105}, 184431 (2022)
\bibitem{xymodel}Das, S., Voleti, S., Saha-Dasgupta, T. and Paramekanti, A. XY magnetism, Kitaev exchange, and long-range frustration in the $J_{eff}= 1/2$ honeycomb cobaltates. {\em Physical Review B}. \textbf{104}, 134425 (2021)
\bibitem{winter2022}Winter, S. Magnetic couplings in edge-sharing high-spin d 7 compounds. {\em Journal Of Physics: Materials}. \textbf{5}, 045003 (2022)
\bibitem{intermediate2024}Mukharjee, P., Shen, B., Erdmann, S., Jesche, A., Kaiser, J., Baral, P., Zaharko, O., Gegenwart, P. and Tsirlin, A. Intermediate field-induced phase of the honeycomb magnet BaCo$_2$(AsO$_4$)$_2$. {\em ArXiv Preprint ArXiv:2403.04466}. (2024)
\bibitem{shiyan}Tu, C., Dai, D., Zhang, X., Zhao, C., Jin, X., Gao, B., Dai, P. and Li, S. Evidence for gapless quantum spin liquid in a honeycomb lattice. {\em ArXiv Preprint ArXiv:2212.07322}. (2022).
\bibitem{THz}Zhang, X., Xu, Y., Halloran, T., Zhong, R., Broholm, C., Cava, R., Drichko, N. and Armitage, N. A magnetic continuum in the cobalt-based honeycomb magnet BaCo$_2$(AsO$_4$)$_2$. {\em Nature Materials}. \textbf{22}, 58-63 (2023).
\bibitem{bose2023proximate}Bose, A., Routh, M., Voleti, S., Saha, S., Kumar, M., Saha-Dasgupta, T. and Paramekanti, A. Proximate Dirac spin liquid in the honeycomb lattice J$_1$-J$_3$ XXZ model: Numerical study and application to cobaltates. {\em Physical Review B}. \textbf{108}, 174422 (2023)

\bibitem{Wolfowicz}Wolfowicz, G., Heremans, F., Anderson, C., Kanai, S., Seo, H., Gali, A., Galli, G. and Awschalom, D. Quantum guidelines for solid-state spin defects. {\em Nature Reviews Materials}. \textbf{6}, 906-925 (2021).
\bibitem{Supplement} See Supplemental Information.
\bibitem{Ponomaryov2017}Ponomaryov, A., Schulze, E., Wosnitza, J., Lampen-Kelley, P., Banerjee, A., Yan, J., Bridges, C., Mandrus, D., Nagler, S., Kolezhuk, A. and Zvyagin, S. Unconventional spin dynamics in the honeycomb-lattice material $\alpha$-RuCl$_3$: High-field electron spin resonance studies. {\em Physical Review B}. \textbf{96}, 241107 (2017).
\bibitem{Ponomaryov2020}Ponomaryov, A., Zviagina, L., Wosnitza, J., Lampen-Kelley, P., Banerjee, A., Yan, J., Bridges, C., Mandrus, D., Nagler, S. and Zvyagin, S. Nature of magnetic excitations in the high-field phase of $\alpha$-RuCl$_3$. {\em Physical Review Letters}. \textbf{125}, 037202 (2020).
\bibitem{Wellm}Wellm, C., Zeisner, J., Alfonsov, A., Wolter, A., Roslova, M., Isaeva, A., Doert, T., Vojta, M., Büchner, B. and Kataev, V. Signatures of low-energy fractionalized excitations in $\alpha$-RuCl$_3$ from field-dependent microwave absorption. {\em Physical Review B}. \textbf{98}, 184408 (2018).
\bibitem{Balz}Balz, C., Lampen-Kelley, P., Banerjee, A., Yan, J., Lu, Z., Hu, X., Yadav, S., Takano, Y., Liu, Y., Tennant, D., Lumsden, M., Mandrus, D. and Nagler, S. Finite field regime for a quantum spin liquid in $\alpha$-RuCl$_3$. {\em Physical Review B}. \textbf{100}, 060405 (2019).
\bibitem{co3v2o8}Zhao, X., Wu, J., Zhao, Z., He, Z., Song, J., Zhao, J., Liu, X., Sun, X. and Li, X. Heat switch effect in an antiferromagnetic insulator Co$_3$V$_2$O$_8$. {\em Applied Physics Letters}. \textbf{108}, 242405 (2016).
\bibitem{nctoHess}Hong, X., Gillig, M., Yao, W., Janssen, L., Kocsis, V., Gass, S., Li, Y., Wolter, A., Büchner, B. and Hess, C. Phonon thermal transport shaped by strong spin-phonon scattering in a Kitaev material Na$_2$Co$_2$TeO$_6$. {\em Npj Quantum Materials}. \textbf{9}, 18 (2024).
\bibitem{rucl3resonance}Hentrich, R., Wolter, A., Zotos, X., Brenig, W., Nowak, D., Isaeva, A., Doert, T., Banerjee, A., Lampen-Kelley, P., Mandrus, D., Nagler, S., Sears, J., Kim, Y., Büchner, B. and Hess, C. Unusual Phonon Heat Transport in $\alpha$-RuCl$_3$: Strong Spin-Phonon Scattering and Field-Induced Spin Gap. {\em Phys. Rev. Lett.}. \textbf{120}, 117204 (2018,3).

\bibitem{tb2ti2o7}Li, Q., Zhao, Z., Fan, C., Zhang, F., Zhou, H., Zhao, X. and Sun, X. Phonon-glass-like behavior of magnetic origin in single-crystal Tb$_2$Ti$_2$O$_7$. {\em Phys. Rev. B}. \textbf{87}, 214408 (2013,6).
\bibitem{homno3}Wang, X., Fan, C., Zhao, Z., Tao, W., Liu, X., Ke, W., Zhao, X. and Sun, X. Large magnetothermal conductivity of HoMnO$_3$ single crystals and its relation to the magnetic-field-induced transitions of magnetic structure. {\em Phys. Rev. B}. \textbf{82}, 094405 (2010,9).

\bibitem{ndmno3}Berggold, K., Baier, J., Meier, D., Mydosh, J., Lorenz, T., Hemberger, J., Balbashov, A., Aliouane, N. and Argyriou, D. Anomalous thermal expansion and strong damping of the thermal conductivity of NdMnO$_3$ and TbMnO$_3$ due to 4f crystal-field excitations. {\em Physical Review B—Condensed Matter And Materials Physics}. \textbf{76}, 094418 (2007).
\bibitem{crcl3}Pocs, C., Leahy, I., Zheng, H., Cao, G., Choi, E., Do, S., Choi, K., Normand, B. and Lee, M. Giant thermal magnetoconductivity in CrCl$_3$ and a general model for spin-phonon scattering. {\em Phys. Rev. Res.}. \textbf{2}, 013059 (2020,1).

\end{thebibliography}
\end{document}